\def\be{\begin{equation}}
\def\ee{\end{equation}}
\def\ba{\begin{array}}

\def\ea{\end{array}}

\documentclass[12pt]{article}
\usepackage{amsfonts,amssymb,amsmath,graphicx}
\newtheorem{theorem}{Theorem}
\newtheorem{corollary}{Corollary}

\begin{document}
\parskip=3pt
\parindent=18pt
\baselineskip=20pt
\setcounter{page}{1}
\centerline{\large\bf Variance-based uncertainty relations for incompatible observables}
\vspace{6ex}
\centerline{{\sf Bin Chen,$^{\star,\dag}$}
\footnote{\sf chenbin5134@163.com}
~~~ {\sf Ning-Ping Cao,$^{\natural}$}
\footnote{\sf sophyc21@163.com}
~~~ {\sf Shao-Ming Fei,$^{\natural,\sharp}$}
\footnote{\sf feishm@cnu.edu.cn}
~~~ {\sf Gui-Lu Long$^{\star,\dag,\ddag}$}
\footnote{\sf gllong@tsinghua.edu.cn}
}
\vspace{4ex}
\centerline
{\it $^\star$ State Key Laboratory of Low-Dimensional Quantum Physics and Department of Physics,}\par
\centerline
{\it Tsinghua University, Beijing 100084, China}\par
\centerline
{\it $^\dag$ Tsinghua National Laboratory for Information Science and Technology, Beijing 100084, China}\par
\centerline
{\it $^\natural$ School of Mathematical Sciences, Capital Normal University, Beijing 100048, China}\par
\centerline
{\it $^\sharp$ Max-Planck-Institute for Mathematics in the Sciences, 04103 Leipzig, Germany}\par
\centerline
{\it $^\ddag$ Collaborative Innovation Center of Quantum Matter, Beijing 100084, China}\par
\vspace{6.5ex}
\parindent=18pt
\parskip=5pt
\begin{center}
\begin{minipage}{5in}
\vspace{3ex}
\centerline{\large Abstract}
\vspace{4ex}
We formulate uncertainty relations for arbitrary finite number of incompatible observables.
Based on the sum of variances of the observables, both Heisenberg-type and Schr\"{o}dinger-type uncertainty relations are provided.
These new lower bounds are stronger in most of the cases than the ones derived
from some existing inequalities. Detailed examples are presented.
\end{minipage}
\end{center}

\newpage

\section{Introduction}
Uncertainty principle is one of the most remarkable features of quantum mechanics.
In 1927, Heisenberg \cite{Heisenberg} introduced the first uncertainty inequality for a pair of canonical observables -- position $x$ and momentum $p$.
This inequality can be expressed in terms of standard deviation-based product uncertainty relation,
$$
\Delta x\Delta p\geq\frac{\hbar}{2},
$$
where the standard deviation of an operator $\Omega$ is defined by $\Delta\Omega=\sqrt{\langle\Omega^{2}\rangle-\langle\Omega\rangle^{2}}$, $\hbar$ is the Planck constant,
$\langle\Omega\rangle$ is the mean value the operator $\Omega$.
After that, Robertson \cite{Robertson} generalized the above inequality to any pair of observables $A$ and $B$, and provided the following uncertainty relation:
\begin{equation}\label{RUR}
\Delta A\Delta B\geq\frac{1}{2}|\langle[A,B]\rangle|,
\end{equation}
where $[A,B]=AB-BA$ is the commutator of $A$ and $B$.
The lower bound of Robertson uncertainty relation (RUR) (\ref{RUR}) has an explicit physical meaning. It can be used to capture the non-commutativity of the two observables.
A strengthened form of RUR is due to Schr\"{o}dinger \cite{Schrodinger}, who derived the following Schr\"{o}dinger uncertainty relation (SUR):
\begin{equation}\label{SUR}
(\Delta A)^{2}(\Delta B)^{2}\geq\left|\frac{1}{2\mathrm{i}}\langle[A,B]\rangle\right|^{2}+\left|\frac{1}{2}\langle\{A,B\}\rangle-\langle A\rangle\langle B\rangle\right|^{2}.
\end{equation}
Uncertainty inequalities of the types of (\ref{RUR}) and (\ref{SUR}) are often referred to as Heisenberg-type and Schr\"{o}dinger-type uncertainty relations, respectively.

Recently, Maccone and Pati \cite{Maccone} provided two stronger uncertainty relations based on the sum of variances.
These uncertainty inequalities are nontrivial whenever the measured state is not a common eigenstate of the two observables.
Thus these new lower bounds of uncertainty relations can capture better the incompatibility of the two observables.
After that, Chen and Fei \cite{Chen} generalized one of the uncertainty relations in Ref. \cite{Maccone} to arbitrary $N$ incompatible observables.

Variance-based uncertainty relations have many useful applications in quantum information theory, such as entanglement detection \cite{Hofmann,Guhne}, quantum spin squeezing (see Ref. \cite{Ma} and references therein), etc.
It is worth noting that there are many other ways to formulate uncertainty relations, such as in terms of entropies \cite{y10,y11,j1,j2,j3,j4}, by use of majorization approach \cite{y12,y13,y131,y132,y14,y141}, based on the skew information \cite{j11,j22,j33} and the notion of fine-grained uncertainty relation \cite{f1,f2,f3}.
Throughout the paper, we only focus on the variance-based sum uncertainty relations.

In this paper, we provide several new uncertainty relations for arbitrary finite number of incompatible observables.
These new lower bounds are shown to be tighter than the ones in the previous works via some examples of spin-$\frac{1}{2}$ particle system.
The paper is organized as follows. In Sect. 2, we present a sum uncertainty relation for $n$ incompatible observables.
When $n=2,3$, the inequality is shown to be stronger than the ones in Ref. \cite{Maccone} and Ref. \cite{Chen} in some cases.
In Sect. 3, we derived a product uncertainty relation, each of the factor in which is the sum of uncertainties in the individual incompatible observables.
We conclude the paper in Sect. 4.

\section{Stronger sum uncertainty relation for $n$ incompatible observables}
In this section, we first provide a stronger sum uncertainty relation for $n$ incompatible observables. We have the following Theorem.

\begin{theorem}
Let $A_{1},A_{2},\ldots,A_{n}$ be $n$ incompatible observables and $\rho$ a quantum state. For each $A_{i}$, if $\Delta A_{i}\neq0$, then we define $P_{i}=\widetilde{A_{i}}\sqrt{\rho}/\Delta A_{i}$, where $\widetilde{A_{i}}=A_{i}-\langle A_{i}\rangle I$, $I$ is the identity operator. If $\Delta A_{i}=0$, we set $P_{i}$ to be a zero operator. Let $M$ be an $n\times n$ matrix with entries $M_{ij}=\mathrm{Tr}(P_{i}^{\dag}P_{j})$. Then we have
\begin{equation}\label{t1}
\sum_{i=1}^{n}(\Delta A_{i})^{2}\geq\frac{1}{\lambda_{\max}(M)}\left[\Delta\left(\sum_{i=1}^{n}A_{i}\right)\right]^{2},
\end{equation}
where $\lambda_{\max}(M)$ denotes the maximal eigenvalue of $M$.
\end{theorem}

\emph{Proof.} We first note that $M$ is a positive semi-definite matrix, since $\sum_{i,j}x_{i}^{*}M_{ij}x_{j}=\|\sum_{i}x_{i}P_{i}\|^{2}\geq0$, $\forall(x_{1},\ldots x_{n})\in\mathbb{C}^{n}$. Then we have $\lambda_{\max}(M)\geq0$.
Taking into account that $\widetilde{\sum_{i}A_{i}}=\sum_{i}\widetilde{A_{i}}$, we have
\begin{eqnarray*}
\left[\Delta\left(\sum_{i}A_{i}\right)\right]^{2} & = & \left\langle\left(\widetilde{\sum_{i}A_{i}}\right)^{2}\right\rangle\\
& = & \left\langle\widetilde{\sum_{i}A_{i}}\widetilde{\sum_{j}A_{j}}\right\rangle\\
& = & \sum_{i,j}\left\langle\widetilde{A_{i}}\widetilde{A_{j}}\right\rangle\\
& = & \sum_{i,j}\Delta A_{i}M_{ij}\Delta A_{j}\\
& \leq & \lambda_{\max}(M)\sum_{i}(\Delta A_{i})^{2}.
\end{eqnarray*}
If $\lambda_{\max}(M)=0$, then $M$ is a zero matrix, and all $\Delta A_{i}=0$, which is contrary to the incompatibility of $A_{i}s$.
Hence $\lambda_{\max}(M)>0$, and $\sum_{i}(\Delta A_{i})^{2}\geq\frac{1}{\lambda_{\max}(M)}[\Delta(\sum_{i}A_{i})]^{2}$ holds from the above inequality. \quad $\Box$

In Ref. \cite{Maccone}, the authors derived an uncertainty relation for two observables $A_{1}$ and $A_{2}$:
\begin{equation}\label{M1}
(\Delta A_{1})^{2}+(\Delta A_{2})^{2}\geq\frac{1}{2}[\Delta(A_{1}+A_{2})]^{2}.
\end{equation}
If $A_{1}$ and $A_{2}$ are incompatible, then the new bound in (\ref{t1}) is shown to be better than the one in (\ref{M1}), which is given in the following corollary.

\begin{corollary}
For two incompatible observables $A_{1}$ and $A_{2}$, we have
\begin{equation}\label{co1}
(\Delta A_{1})^{2}+(\Delta A_{2})^{2}\geq\frac{1}{\lambda_{\max}(M)}[\Delta(A_{1}+A_{2})]^{2}\geq\frac{1}{2}[\Delta(A_{1}+A_{2})]^{2},
\end{equation}
where $M$ is defined in Theorem 1.
\end{corollary}

\emph{Proof.} We only need to prove that $\lambda_{\max}(M)\leq2$. If $\Delta A_{1}=0$ or $\Delta A_{2}=0$, then $\lambda_{\max}(M)=1$.
Suppose that $\Delta A_{1}\neq0$ and $\Delta A_{2}\neq0$. Then $\lambda_{\max}(M)=1+|\mathrm{Tr}(P_{1}^{\dag}P_{2})|$.
Note that $\|P_{1}\|=\|P_{2}\|=1$, we have $|\mathrm{Tr}(P_{1}^{\dag}P_{2})|\leq1$ by the Cauchy-Schwarz inequality. Hence $\lambda_{\max}(M)\leq2$ and the last inequality in (\ref{co1}) holds. \quad $\Box$

\begin{figure}
\centering
\includegraphics[width=7cm]{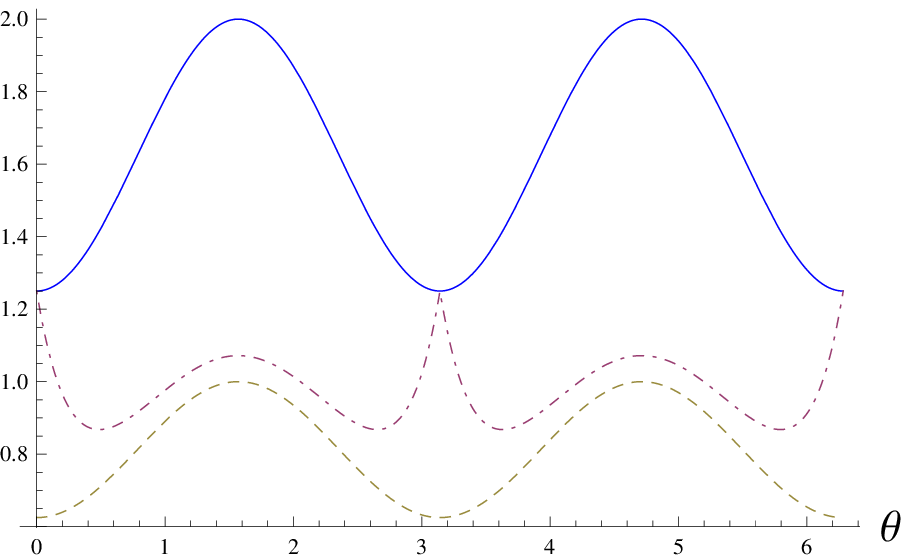}
\caption{The blue solid line is the sum of the variances $(\Delta\sigma_{1})^{2}+(\Delta\sigma_{3})^{2}$.
The dot-dashed line is the bound (\ref{t1}). The dashed line is the bound (\ref{M1}).}\label{cb91}
\end{figure}

To see that our new lower bound (\ref{t1}) is strictly greater than the one in (\ref{M1}), consider the standard Pauli matrices $\sigma_{1}$ and $\sigma_{3}$. Let the qubit state to be measured
and the measured state given by the Bloch vector $\overrightarrow{r}=(\frac{\sqrt{3}}{2}\cos\theta,\frac{\sqrt{3}}{2}\sin\theta,0)$. Then we have $\Delta\sigma_{1}=\sqrt{1-\frac{3}{4}\cos^{2}\theta}$, $\Delta\sigma_{3}=1$, $\Delta(\sigma_{1}+\sigma_{3})=\sqrt{2-\frac{3}{4}\cos^{2}\theta}$,
$\lambda_{\max}(M)=1+\frac{\sqrt{3}|\sin\theta|}{\sqrt{1+3\sin^{2}\theta}}$.
The comparison between the lower bounds (\ref{t1}) and (\ref{M1}) is shown in Fig. \ref{cb91}. Apparently our new bound is strictly greater than (\ref{M1}).

Recently, Chen and Fei \cite{Chen} generalize uncertainty relation (\ref{M1}) to the case of arbitrary $n$ incompatible observables, and present the following
variance-based sum uncertainty relation:
\begin{equation}\label{cf}
\begin{split}
\sum_{i=1}^{n}(\Delta A_{i})^{2}\geq&\frac{1}{n-2}\left\{\sum_{1\leq i<j\leq n}[\Delta(A_{i}+A_{j})]^{2}-\frac{1}{(n-1)^{2}}\left[\sum_{1\leq i<j\leq n}\Delta(A_{i}+A_{j})\right]^{2}\right\}.
\end{split}
\end{equation}

\begin{figure}
\centering
\includegraphics[width=7cm]{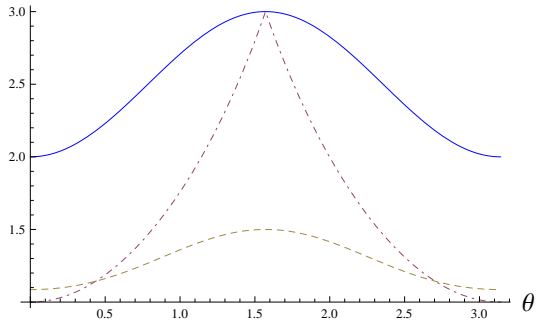}
\caption{The blue solid line is the sum of the variances $(\Delta\sigma_{1})^{2}+(\Delta\sigma_{2})^{2}+(\Delta\sigma_{3})^{2}$.
The dot-dashed line is the bound (\ref{t1}). The dashed line is the bound (\ref{cf}).}\label{cb92}
\end{figure}

To compare the uncertainty relation (\ref{t1}) with (\ref{cf}), let us consider again the Pauli matrices $\sigma_{1},\sigma_{2},\sigma_{3}$, and the measured state given by the Bloch vector $\overrightarrow{r}=(\cos\theta,0,0),\theta\in(0,\pi)$. Then we have
$(\Delta\sigma_{1})^{2}+(\Delta\sigma_{2})^{2}+(\Delta\sigma_{3})^{2}=3-\cos^{2}\theta$,
$\Delta(\sigma_{1}+\sigma_{2})=\Delta(\sigma_{1}+\sigma_{3})=\sqrt{2-\cos^{2}\theta}$, $\Delta(\sigma_{2}+\sigma_{3})=\sqrt{2}$,
$\Delta(\sigma_{1}+\sigma_{2}+\sigma_{3})=\sqrt{3-\cos^{2}\theta}$, $\lambda_{\max}(M)=1+|\cos\theta|$.
It is shown in Fig. \ref{cb92} that for a wide range of $\theta$, our new uncertainty relation (\ref{t1}) is stronger than (\ref{cf}).

\section{Uncertainty relation based on the product of sum of variances in the individual observables}
In Ref. \cite{Pati}, the author studied the product of sum of uncertainties in the individual observables.
They investigated a special case for a set of non-commuting observables.
Using the RUR (\ref{RUR}) and the convexity of quantum uncertainty, they formulated the product of sum of uncertainties in $\{A_{i}\}_{i=1}^{n}$ and $\{B_{i}\}_{i=1}^{n}$, which satisfy $[A_{i},B_{j}]=\mathrm{i}\delta_{ij}C$, and presented the following uncertainty relation:
$$
\left(\sum_{i}^{n}\Delta A_{i}\right)\left(\sum_{i}^{n}\Delta B_{i}\right)\geq\frac{n}{2}|\langle C\rangle|.
$$
Here we generalize the idea to any two sets of incompatible observables. We have the following Schr\"{o}dinger-type uncertainty relation.

\begin{theorem}
Let $\{A_{i}\}_{i=1}^{n}$ and $\{B_{j}\}_{j=1}^{m}$ be two sets of incompatible observables and $\rho$ a quantum state.
If $\Delta A_{i}\neq0$, then we define $X_{i}=\widetilde{A_{i}}\sqrt{\rho}/\Delta A_{i}$, otherwise define $X_{i}=O$;
similarly, if $\Delta B_{j}\neq0$, we define $Y_{j}=\widetilde{B_{j}}\sqrt{\rho}/\Delta B_{j}$, otherwise $Y_{j}=O, \forall i,j$.
Let $G$ be an $n\times m$ matrix with entries $G_{ij}=|\mathrm{Tr}(X_{i}^{\dag}Y_{j})|$.
If there exist $A_{i},B_{j}$ such that $\langle \widetilde{A_{i}}\widetilde{B_{j}}\rangle\neq0$, then we have
\begin{equation}\label{t2}
\begin{split}
\left[\sum_{i=1}^{n}(\Delta A_{i})^{2}\right]^{\frac{1}{2}}\left[\sum_{j=1}^{m}(\Delta B_{j})^{2}\right]^{\frac{1}{2}}
\geq&\frac{1}{\sigma_{\max}(G)}\sum_{i,j} \left(\left|\frac{1}{2\mathrm{i}}\langle[A_{i},B_{j}]\rangle\right|^{2}\right.\\
&\left.+\left|\frac{1}{2}\langle\{A_{i},B_{j}\}\rangle-\langle A_{i}\rangle\langle B_{j}\rangle\right|^{2}\right)^{\frac{1}{2}},
\end{split}
\end{equation}
where $\sigma_{\max}(G)$ is the maximal singular value of $G$.
\end{theorem}

\emph{Proof.} We first have
\begin{eqnarray*}
\sum_{i,j}|\langle \widetilde{A_{i}}\widetilde{B_{j}}\rangle| & = & \sum_{i,j}\Delta A_{i}G_{ij}\Delta B_{j}\\
& \leq & \sigma_{\max}(G)\left[\sum_{i=1}^{n}(\Delta A_{i})^{2}\right]^{\frac{1}{2}}\left[\sum_{j=1}^{m}(\Delta B_{j})^{2}\right]^{\frac{1}{2}}.
\end{eqnarray*}
On the other hand,
\begin{eqnarray*}
|\langle \widetilde{A_{i}}\widetilde{B_{j}}\rangle| & = & |\langle A_{i}B_{j}\rangle-\langle A_{i}\rangle\langle B_{j}\rangle|\\
& = & \left(\left|\frac{1}{2\mathrm{i}}\langle[A_{i},B_{j}]\rangle\right|^{2}
+\left|\frac{1}{2}\langle\{A_{i},B_{j}\}\rangle-\langle A_{i}\rangle\langle B_{j}\rangle\right|^{2}\right)^{\frac{1}{2}}.
\end{eqnarray*}
If $\sigma_{\max}(G)=0$, then all $\langle \widetilde{A_{i}}\widetilde{B_{j}}\rangle=0$, which is contrary to our assumption.
Taking into account the above formulas, we get (\ref{t2}) directly. \quad $\Box$

\emph{Remark}. It is reasonable to assume that not all $\langle \widetilde{A_{i}}\widetilde{B_{j}}\rangle=0$, since otherwise we have
$\sum_{i}(\Delta A_{i})^{2}\sum_{j}(\Delta B_{j})^{2}\geq\sum_{i,j}|\langle \widetilde{A_{i}}\widetilde{B_{j}}\rangle|^{2}=0$
by SUR (\ref{SUR}), which gives rise to a trivial uncertainty relation.

We can also obtain the following Heisenberg-type uncertainty relation from (\ref{t2}).

\begin{corollary}
Under the conditions of the Theorem 2, we have
\begin{equation}\label{c2}
\left[\sum_{i=1}^{n}(\Delta A_{i})^{2}\right]^{\frac{1}{2}}\left[\sum_{j=1}^{m}(\Delta B_{j})^{2}\right]^{\frac{1}{2}}
\geq\frac{1}{2\sigma_{\max}(G)}\sum_{i,j}|\langle[A_{i},B_{j}]\rangle|.
\end{equation}
\end{corollary}

\begin{figure}
\centering
\includegraphics[width=7cm]{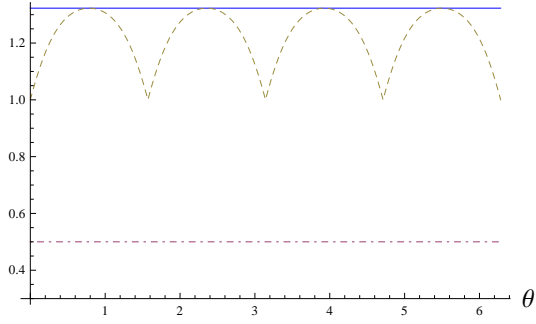}
\caption{The blue solid line is the uncertainties $\Delta\sigma_{3}\sqrt{(\Delta\sigma_{2})^{2}+(\Delta\sigma_{1})^{2}}$.
The dot-dashed line is the bound (\ref{c22}). The dashed line is the bound (\ref{c2}).}\label{cb93}
\end{figure}

Noting that the following uncertainty relation holds by RUR (\ref{RUR}),
\begin{equation}\label{c22}
\left[\sum_{i=1}^{n}(\Delta A_{i})^{2}\right]^{\frac{1}{2}}\left[\sum_{j=1}^{m}(\Delta B_{j})^{2}\right]^{\frac{1}{2}}
\geq\frac{1}{2}\left(\sum_{i,j}|\langle[A_{i},B_{j}]\rangle|^{2}\right)^{\frac{1}{2}},
\end{equation}
we need to compare the lower bounds (\ref{c2}) and (\ref{c22}). Let $A_{1}=\sigma_{3},B_{1}=\sigma_{1},B_{2}=\sigma_{2}$, and the Bloch vector of the measured state $\overrightarrow{r}=(\frac{1}{2}\cos\theta,\frac{1}{2}\sin\theta,0)$. It is shown in Fig. \ref{cb93} that the uncertainty relation (\ref{c2}) we derived in this paper is stronger in this case than the one trivially obtained by summing over RURs for all pairs of $A_{i}$ and $B_{j}$.

Let us consider a special case for $n=m$ and $A_{i}=B_{i}$. We have the following uncertainty relation.

\begin{corollary}
Let $A_{1},A_{2},\ldots,A_{n}$ be $n$ incompatible observables. We have
\begin{equation}
\sum_{i=1}^{n}(\Delta A_{i})^{2}\geq\frac{1}{\min\{\sigma_{\max}(G),n-1\}}\sum_{1\leq i<j\leq n}|\langle[A_{i},A_{j}]\rangle|,
\end{equation}
where $G$ is defined in Theorem 2.
\end{corollary}

\emph{Proof.} By using RUR (\ref{RUR}) and the mean inequality for pairs of observables $A_{1},\ldots,A_{n}$, we get
$$
(\Delta A_{i})^{2}+(\Delta A_{j})^{2}\geq2\Delta A_{i}\Delta A_{j}\geq|\langle[A_{i},A_{j}]\rangle|, ~~ \forall i\neq j.
$$
Summing over the above inequalities, we have
\begin{equation}
\sum_{i=1}^{n}(\Delta A_{i})^{2}\geq\frac{1}{n-1}\sum_{1\leq i<j\leq n}|\langle[A_{i},A_{j}]\rangle|.
\end{equation}

We now only need to show that $\sigma_{\max}(G)<n-1$ for some $n$, when suitable observables and measured state considered.
Let $n=3$ and $A_{i}=\sigma_{i}$, $i=1,2,3$, $\rho$ is given by the Bloch vector $\overrightarrow{r}=(\cos\theta,0,0),\theta\in(0,\pi)$.
Then we have $\sigma_{\max}(G)=1+|\cos\theta|<2$. This completes the proof. \quad $\Box$

\section{Conclusion}
We have formulated uncertainty relations for arbitrary incompatible observables. These new uncertainty inequalities are based on the sum of variances of the observables. The corresponding lower bounds we derived in this paper are shown to be tighter than the previous ones. Thus our bounds capture better the incompatibility of the observables. We have also studied the general form of product of sum of variances in the individual observables, and obtained a stronger uncertainty relation.
We emphasize that these new bounds are physically realizable in general. These bounds are tight and can be reached by detailed quantum systems, as shown in examples. Moreover, they are also physically measurable. For the uncertainty relation (\ref{t1}), we only need to measure the variance of sum of the observables, and the value of $\lambda_{\max}(M)$ can be estimated beforehand. Similarly, for the uncertainty relation (\ref{t2}), we need to measure the mean value of each observable and the mean value of commutators and anti-commutators of pairs of the observables. The value of $\sigma_{\max}(G)$ can be computed readily.
More interesting results could be obtained if multipartite systems are taken into account. As uncertainty relations are tightly related to many quantum tasks, our results might shed new lights on investigating quantum information processing like entanglement detection, quantum spin squeezing, quantum separability criteria, security analysis of quantum key distribution in quantum cryptography and nonlocality.

\vspace{2.5ex}
\noindent{\bf Acknowledgments}\, \,
This work is supported by the National Basic
Research Program of China (2015CB921002), the National Natural Science Foundation
of China Grant Nos. 11175094, 91221205 and 11275131.

\end{document}